\documentclass[preprint,aps,showpacs]{revtex4}
\usepackage{amsfonts}
\usepackage{amsmath}
\usepackage{amssymb}
\usepackage{graphicx}
\usepackage{epsfig}

\providecommand{\U}[1]{\protect\rule{.1in}{.1in}}
\providecommand{\U}[1]{\protect\rule{.1in}{.1in}}
\providecommand{\U}[1]{\protect\rule{.1in}{.1in}}
\providecommand{\U}[1]{\protect\rule{.1in}{.1in}}

\input{tcilatex}
\begin{document}

\title[ ]{Evolving Network With Different Edges}
\author{Jie Sun, Yizhi Ge and Sheng Li$^{\ast}$}
\affiliation{Department of Physics, Shanghai Jiao Tong University, Shanghai, China}
\pacs{84.35.+i, 05.40.-a, 05.50.+q,
87.18.Sn}

\begin{abstract}
We proposed an evolving network model constituted by the same nodes but
different edges. The competition between nodes and different links were
introduced. Scale free properties have been found in this model by continuum
theory. Different network topologies can be generated by some tunable
parameters. Simulation results consolidate the prediction.
\end{abstract}

\maketitle

\section{Introduction}

For the past few years, the interesting properties of complex network, which
can be found in many fields, such as biology, sociology, physics, and so on,
have attracted a number of people to explore. Great contribution including
the idea and analysis of random networks\cite{Erods}, small world networks
such as the WS network\cite{Watts1,Watts2,Newman} has been made to the
modeling process\cite{Erods,Derrida,Derrida2,Monasson}. However, real world
networks such as Internet, Movie actor and science collaboration graph(see
\cite{Watts1,Amaral} all show a power-law degree distribution property,
which can not be explained by the two models mentioned above. A scale free
model\cite{Barabasi,Barabasi2} that satisfies a lot of real systems and led
many subsequent researchers to further studying has been found by Barabasi
and Albert in 1999. On the other hand, this model suffers a drawback that
the exponent of the power law is always fixed, while in real networks it
varies. Further study has made it explicit that preferential attachment
plays a key role in deciding whether the evolving network shows scaling
property as expected\cite%
{Barabasi,Barabasi2,Dorogovtsev1,Dorogovtsev2,Dorogovtsev3}. In 2000, an
exact solution to the degree distribution of BA model was obtained by
Dorogovtsev, Mendes and Samukhin\cite{Dorogovtsev3}. In their paper,
different initial attractiveness is used, to change the exponent of the
power-law curve.

Network models are applied to the description of the complex human
interactions, known as social networks. Enlightened by the fact that there
are complicated relationships between the elements of social networks,
researchers brought multiple-type vertex rather than single one, into
network models\cite{kim}. Weighted edges as represented by kinds of social
relationships are also introduced in this area\cite{MB}. Both models show
scaling properties. It is challenging to find more fundamental network
structures.

Similarly, our model can be understood from another perspective. One way,
but not the exclusive one, to understand our concept is to consider the
Internet as a network, which comprised by online websites as nodes. We set
the virtual connections, e.g. the transportation forms of information
between different websites rather than physical wires and cables, as edges
between the nodes. It is clear that there are various kinds of communication
ways between two websites( HTTP, FTP ). So to speak, different kinds of
links need to be applied to growing networks, where competitions exist
between different links as well as the nodes. It is attractive to discover
the topologies of the whole network and different sub-networks containing
only one kind of the two edges, and to consider the relation between
different edges. In the model we construct, the whole network is somewhat
like the partial preference BA model. The significant difference is that we
add some tunable parameters into the preferential attachment, which makes it
possible to alter the exponent of the power-law degree distribution and the
shift constant of degree distribution function. Our work begins with the
discussion of a scale free network growing like the BA model with a change
of the preferential attachment. This is a special example of the growing
networks put forward by Dorogovstev and Mendes in 2000\cite{Dorogovtsev3} in
which they add a constant in the preferential attachment term. In our main
work, the tunable parameters in the model play different roles in different
edge preferential attachments, and are regarded as a weight to discriminate
different attractiveness of different links in the evolving network. The
research produces some interesting results: the whole network and the two
sub-networks all evolve just like the ordinary scale free network, while the
two kinds of links connected to a certain node always differ.\ Besides, the
degree distributions of the two sub-networks are unequal.

\section{Model and theoretical approach}

At the beginning, we revise the BA model only a bit and get different
expressions of the degree distribution function. Before introducing our
findings, it is necessary to take a look at the work by Dorogovstev and
Mendes in 2000\cite{Dorogovtsev3}. Their contribution mainly lies with the
modulation of the original BA model and the way they solved the problem,
well known as the 'Master Equation Approach'. In their paper, the
preferential attachment in BA model is no longer fixed; instead, changeable
initial attractiveness, which is given to each site, together with incoming
links of the site, determines the probability whether a new link will point
to this site. Under long-time limitation, the exponent of the degree
distribution varies from $2$ to $\infty $ , depending on the initial
attractiveness. In this paper we do not intend to discuss too many
complicated forms of preferential attachments deriving from the BA model. We
barely make a simple and clear revision, and attain concise and
approximately precise results, with the application of continuum theory.

The preferential attachment in BA model is changed in this way: the
possibility of the degree increase of a site is proportional to $\prod
(k_{i})=m\frac{k_{i}+f}{\sum (k_{i}+f)}$($k_{i}$ is the degree of site i,
while $f$ is a given constant). This model may be considered as a special
example of the model generalized by Dorogovestev and Mendes if one takes $%
A^{(0)}$ in their model as $m+f$ in our case. By making use of master
equation to solve this problem, the degree distribution was found to have
the exponent $-(2+a)$ where $a=\frac{A^{(0)}}{m}$. A linear shift $%
ma=A^{(0)} $ is found in the distribution function
\begin{equation}
P(q)\approx (1+a)\frac{\Gamma \lbrack (m+1)a+1]}{\Gamma (ma)}(q+ma)^{-(2+a)}
\label{pk1}
\end{equation}%
We here apply continuum theory, rather than master equation, to solve the
problem we raise. It is not difficult to draw the following equation:%
\begin{equation}
\frac{\partial k_{i}}{\partial t}=m\frac{k_{i}+f}{\sum (k_{i}+f)}
\end{equation}%
where $m$ is the number of links every time the network increases. The
solution is quite simple: as $t$ approaches infinity, the degree
distribution follows this equation
\begin{equation}
P(k)=(2+\frac{f}{m})(m+f)^{2+\frac{f}{m}}(f+k)^{-(3+\frac{f}{m})}
\label{pk2}
\end{equation}%
Compare the degree functions (\ref{pk1}) and (\ref{pk2}), taking the power
term into account, and we find that the two results are the same in essence.
In our derivation the tunable parameter is $f$, while in D\&M's it's the
initial attractiveness $A^{(0)}$. By transforming $A^{(0)}$ to $m+f$ or
inversely, the two functions show the same property. This transformation is
reasonable. If we adopt D\&M's way\cite{Dorogovtsev3} in our model, $A^{(0)}$
should be the initial degree a site has when it enters the network, which is
exactly $m+f$. In D\&M's paper\cite{Dorogovtsev3}, the exponent $\gamma =2+a$
varies from $2$ to $\infty $ while in our model, $\gamma =3+\frac{f}{m}$. $f$
can be taken from $-m$ to $\infty $. Therefore, the range is the same. The
advantage by using continuum theory is that no special function appears,
which makes it relatively easy to solve the two-link networks we conceive in
the following content. Another difference between our model and D\&M's model
is that we consider the network as an undirected network while they consider
the network as directed.

The following is our new proposal based on the idea of a new kind of
network. This network, which contains two kinds of edges, is very different
from the networks we have studied till now. If we consider the different
links as identical, the model should come back to partial preferential
attachment BA case. To reach this point briefly, we give different linear
preferential attachments to two kinds of edges respectively. Of course,
modulatory parameters are added to make our model changeable to fit the real
networks.

\emph{The model: }At each time interval, a new node adds to the network and
connects to the old nodes with $m$ edges. We divide the $m$ edges into two
types: one is called $X$ edge(or $X$ link) and the other $Y$ edge(or $Y$
link). The probability that node $i$ in the original network makes a
connection to the new added node follows such preferential attachments:%
\begin{equation}
\dfrac{\partial x_{i}}{\partial t}=m\dfrac{x_{i}+y_{i}+f+gy_{i}}{2(\sum
x_{j}+y_{j}+f)}  \label{ex}
\end{equation}%
\begin{equation}
\dfrac{\partial y_{i}}{\partial t}=m\dfrac{x_{i}+y_{i}+f-gy_{i}}{2(\sum
x_{j}+y_{j}+f)}  \label{ey}
\end{equation}%
where $x_{i}$ represents the number of $X$-edges node $i$ has connected at
time $t$; $y_{i}$ represents the number of $Y$-edges.

Here, we give linear asymmetric expressions in $X$ and $Y$ edge aiming to
represent the linear relations in two networks. However, we do not assume
this to be exactly the reflection of real world networks, it is a step
forward in the direction of finding out complex relations between networks
that share same nodes but different links. The parameter $g$ here plays the
key role in determining the relation between $X$ and $Y$ network: When $g$
is very small, it can be neglected so that the two networks evolve in the
same way-- they just behave as the same network. When $g$ approaches
infinity, say, some relatively large number, the above expressions show that
new edges will always be added as $X$ edges while the number of $Y$ edges
could only remain constant instead of increasing. In general, $g$ is used to
determine the difference of the increasing rates of the two different edges.

The number of total edges node $i$ connects satisfies the following equation:%
\begin{equation}
\dfrac{\partial (x_{i}+y_{i})}{\partial t}=m\dfrac{x_{i}+y_{i}+f}{\sum
x_{j}+y_{j}+f}
\end{equation}%
We have studied this kind of network before, $x_{i}+y_{i}=k_{i}$ and degree
distribution follows equation (\ref{pk2}).

Firstly, we discuss $Y$ sub-network. Because we add $m$ edges at every time
step, so%
\begin{equation}
\sum (x_{i}+y_{i}+f)=(2m+f)t
\end{equation}%
According to continuum theory, equation (\ref{ey}) became%
\begin{equation}
\dfrac{\partial y_{i}}{\partial t}=m\frac{(m+f)(\frac{t}{t_{i}})^{\frac{m}{%
2m+f}}-gy_{i}}{2(2m+f)t}
\end{equation}%
Its solution is%
\begin{equation}
y_{i}=\frac{m+f}{2+g}(\frac{t}{t_{i}})^{\frac{m}{2m+f}}+Const\cdot t^{-(%
\frac{mg}{2(2m+f)})}  \label{gg}
\end{equation}%
When $t$ approaches infinity, $t^{-(\frac{mg}{2(2m+f)})}$ approaches $0$ and
$y_{i}$ becomes%
\begin{equation}
y_{i}\approx \frac{m+f}{2+g}(\frac{t}{t_{i}})^{\frac{m}{2m+f}}  \label{yi}
\end{equation}%
Then we obtain

\begin{align}
P_{y}(y_{i}& <y)=P(t_{i}>y^{-(2+\frac{f}{m})}(\frac{m+f}{2+g})^{(2+\frac{f}{m%
})}t))  \notag \\
& =1-\frac{1}{m_{0}+t}y^{-(2+\frac{f}{m})}(\frac{m+f}{2+g})^{(2+\frac{f}{m}%
)}t
\end{align}%
For large $t$ ,$P(y_{i})$ reads%
\begin{equation}
P_{y}(y)=\frac{\partial P(y_{i}<y)}{\partial y}\approx \left( 2+\frac{f}{m}%
\right) (\frac{m+f}{2+g})^{(2+\frac{f}{m})}y^{-(3+\frac{f}{m})}  \label{pyr}
\end{equation}%
Up to now, we have got the mathematical expression of degree distribution of
$Y$ sub-network. i.e. equation (\ref{pyr}). The degree distribution is power
law with the same exponent as that of total degree distribution but without
shift $f$. Similarly, we proceed in $X$ sub-network. According to equation (%
\ref{ex}) and (\ref{yi}), we get%
\begin{align}
x_{i}& =k_{i}-y_{i}  \notag \\
& =\frac{1+g}{2+g}(m+f)(\frac{t}{t_{i}})^{(\frac{m}{2m+f})}-f-Const\cdot
t^{-(\frac{mg}{2(2m+f)})}
\end{align}%
Under long-time limitation, we neglect $Const\cdot t^{-(\frac{mg}{2(2m+f)})}$%
, and obtain%
\begin{equation}
x_{i}=\frac{1+g}{2+g}(m+f)(\frac{t}{t_{i}})^{(\frac{m}{2m+f})}-f
\end{equation}%
The same as what we dealt with $Y$ sub-network, the degree distribution of $%
X $ sub-network becomes%
\begin{equation}
P_{x}(x)=\left( 2+\frac{f}{m}\right) [\frac{1+g}{2+g}(m+f)]^{(2+\frac{f}{m}%
)}(x+f)^{-(3+\frac{f}{m})}
\end{equation}%
It is a power-law distribution with shift $f$. The preferential attachments (%
\ref{ex}) and (\ref{ey}) are not the same. Therefore, the number of $x$
edges does not equal to the number of $y$ edges. From the calculations
above, we get the ratios of total degree to $x$ degree and $y$ degree
\begin{align}
P_{x+y}\left( k-f\right) & :P_{x}\left( k-f\right) :P_{y}\left( k\right)
\notag \\
& =1:\left( \frac{1+g}{2+g}\right) ^{2+\frac{f}{m}}:\left( \frac{1}{2+g}%
\right) ^{2+\frac{f}{m}}  \label{ratio}
\end{align}%
the subscript $x+y$ in $P_{x+y}\left( k\right) $ means the degree in total
network.

\section{Numerical simulations}

We made computer simulations of the model. The results accorded with the
theoretical predictions quite well at most parameters ranges. Figure \ref%
{figure1}(a), (b) and (c) give simulations of networks with size $N=1000000$%
, $m=5$ and $f=-2,2,5$ respectively. With log-log scale, the main part of
the distributions give slope lines in the figure. The slopes of linear fit
approximately equal to the theoretical results ($\gamma =2.6,3.4,4$
respectively). The prediction of the ratios of $x$ degree to $y$ degree are
also confirmed by numerical simulations (see figure \ref{figure2}). One
should notice that in the figures, we had made shifts for $x$ degree
distributions and total degree distributions from $P_{x}\left( k\right) $
and $P_{x+y}\left( k\right) $ to $P_{x}\left( k-f\right) $ and $%
P_{x+y}\left( k-f\right) $ respectively. Therefore, their log-log plots
behave as straight lines parallel to those of $y$ degree.

However, we found that when we take small $g$, the simulations are always
different from theoretical predictions. With small $g$, when $f$ is
positive, the component of $x$ degree is larger than theoretical value while
the component $y$ degree accords with theoretical value (see figure \ref%
{figure3}(a)); and when $f$ is negative, the component of $y$ degree is
larger than theoretical value while the component of $x$ degree accords with
theoretical value(see figure \ref{figure3}(b)). In theoretical calculus, at
infinite $t$ limitation, we neglect the last term of equation (\ref{gg}).
When $g$ is small, decay of this term is slow, and actually in numerical
simulations, $t$ is not infinite. Therefore, the neglect of this term make
the bias of simulation results from predictions at small $g$.

We also looked at the fluctuation of $x$ degree $x_{k}$ to theoretical value
as a function of total degree $k$. It illustrates how the competition
generates heterogeneity in edge composition of each vertex. We calculated
the relative standard deviation respectively to theoretical value. The
fluctuation decreases quickly with the increase of total degree nearly in a
power law way (see figure \ref{figure4}). We did not calculate the standard deviation of $%
x_{k}/y_{k}$ because $y$ degree may be zero for some vertexes.

\section{Discussions}

The mathematical expressions of degree distribution of both $Y$ sub-network
and $X$ sub-network provide deep insight into the dynamics of evolving
systems. We build a competitive environment where not only nodes but also
different types of links compete. This model can reflect many properties of
social network. A newly added node has a fixed number of $m$ adding edges.
However, other nodes decided how many $X$ edges and $Y$ edges there would
be. Obviously, other nodes compete for these $m$ edges by what they already
have. For instance, let $X$-edge denotes financial relationship between
individual persons, and $Y$ represents other connection. Rich people tend to
have more financial relationship with other people, while interestingly,
more financial links signifies his richness. But when we focus on other
links between people, rich people are not necessarily to be so lucky. Due to
the limitation of personal capability, time, devotion, one can not have
infinite connections with others. Therefore, it is at the expense of less $Y$
edges to get more $X$ edges, and vice versa. The above study shows the
relationship between $X$ and $Y$ edges.

An important character of this model is that the degree distribution of $X$
sub-network shows a linear shift while $Y$ sub-network does not. Simulations
confirm this character, though the exact value of the shift may has some
errors, due to the application of mean-field approximation. The difference
of linear shift between $X$ and $Y$ sub-network comes from different
preferential attachment.

In the following, we will discuss how the variables $g,$ $f$ and $m$ work on
in our theory.

$m$ is always thought as a fixed number in a certain network, for we can
identify at least vaguely how many links are added in each time interval.
The value of $f$ varies from $-m$ to $\infty $, but if $f$ is too large both
of total degree distribution and $x$ degree distribution approach
exponential increment,%
\begin{align}
(k+f)^{-(3+\frac{f}{m})}& =e^{-(3+\frac{f}{m})\ln (k+f)}=e^{-(3+\frac{f}{m}%
)[\ln f+\ln (1+\frac{k}{f})]}  \notag \\
& \approx f^{-(3+\frac{f}{m})}e^{-(3+\frac{f}{m})\frac{x}{f}}=f^{-(3+\frac{f%
}{m})}e^{-x(\frac{1}{m}+\frac{3}{f})}
\end{align}%
And if $f>>m$, the characteristic degree of total network and $X$
sub-network is $m$. On the other hand, when $f$ is taken a bit large, the
actual probability to connect $Y$ links will be so small that in the end
most of the nodes have small $y$ degree, while the rest have very large $y$
degree.

Note that in the several derivations above, we neglect the term $Const\cdot
t^{-(\frac{mg}{2(2m+f)})}$ for time $t$ is considered approaching to
infinite. However, this ideal condition can not be reached to apply the
model, i.e. $t$ will always be finite actually, even considerable large. So
problem rises that if $g,$ $f$ and $m$ are chosen or set so that the
component $\frac{mg}{2(2m+f)}$ is small enough and the whole term becomes an
unneglectable factor to the degree distribution. The derivation above may
not hold true. But if we take this term into account, we will find ourself
awkward in searching a precise solution for the function of degree
distribution.

The most intriguing parameter in the parameters is $g$. Apparently, from
above discussion, $g$ should not be chosen too small to make the term $t^{-(%
\frac{mg}{2(2m+f)})}$ unneglectable. The simulations showed that small $g$
make the degree distributions depart from predictions. $g$ should not be too
large, as well. A very large $g$ in finite time lead the network to have few
$Y$ links also, as we have found by computer simulation. Based on above
reasons, $g$ is preferred to be such a number that is neither too large, nor
too small. Here we only use vague words "large" and "small", for these
limitations are due to the finity of time a network evolves and the finity
of nodes and links it contains. To say a bit more accurate, we need $g$ to
be big enough, with a view to $t$, to make the term $t^{-(\frac{mg}{2(2m+f)}%
)}$ small enough to be neglected; but $g$ should not be too larger to make $%
\dfrac{\partial x_{i}}{\partial t}=m\dfrac{x_{i}+y_{i}+f+gy_{i}}{2(\sum
x_{j}+y_{j}+f)}$ too larger than $\dfrac{\partial y_{i}}{\partial t}=m\dfrac{%
x_{i}+y_{i}+f-gy_{i}}{2(\sum x_{j}+y_{j}+f)}$, which leads the network
having few $Y$ edges.

For a given node with degree $k$ at given time, the probability
$p_{x}\left( k\right) $ that how many $X$ edges it has follows
equation (\ref{ex}). The figure \ref{figure4} shows the
fluctuation decreases quickly with $k$. It seems that the
distribution of fluctuation should follows the binomial
distribution.
However, here the degree of a node is varying with time. It increased from $%
m $ to $k$ and $p_{x}$ varied synchronously. It results in the fact that the
fluctuation of the number of $X$ edges is much smaller than that of the
binomial distribution.

Finally we discuss directed network and find some difference between this
kind of network and what we have studied. We give general results below.
Every time step, we introduce $m$ directed edges. The whole degree has a $m$
increment (see \cite{Dorogovtsev3}).
\begin{equation*}
\sum \left( k_{i}+f\right) =\left( m+f\right) t
\end{equation*}%
using the same approaches, we get%
\begin{equation}
P_{x+y}(k)=(1+\frac{f}{m})f^{1+\frac{f}{m}}(k+f)^{-(2+\frac{f}{m})}
\label{t1}
\end{equation}%
And the degree distributions of $X$ subnetwork and $Y$ subnetwork are%
\begin{equation}
P_{x}(k)=\left( 1+\frac{f}{m}\right) [\frac{f(2+g)-1}{2+g}]^{(1+\frac{f}{m}%
)}(k+f)^{-(2+\frac{f}{m})}
\end{equation}%
\begin{equation}
P_{y}(k)=\left( 1+\frac{f}{m}\right) (\frac{1}{2+g})^{(1+\frac{f}{m})}k^{-(2+%
\frac{f}{m})}
\end{equation}%
The ratios of total degree to $X$ degree and $Y$ degree are%
\begin{align*}
P_{x+y}\left( k-f\right) & :P_{x}\left( k-f\right) :P_{y}\left( k\right) \\
& =1:\left( \frac{f(2+g)-1}{f(2+g)}\right) ^{1+\frac{f}{m}}:\left( \frac{1}{%
f(2+g)}\right) ^{1+\frac{f}{m}}
\end{align*}%
Compared with the results of undirected network (\ref{ratio}), one
can find that proportion of $Y$ edges of directed network is
larger than that of undirected network.

\begin{acknowledgments}
This paper was supported by the National Science Foundation of China under
Grant No. 10105007, and PRP Foundation of Shanghai Jiao Tong University
Grant No. T0720701.
\end{acknowledgments}

$^{\ast}$ Email: lisheng@sjtu.edu.cn

\begin{figure}[tbp]
\epsfig{file=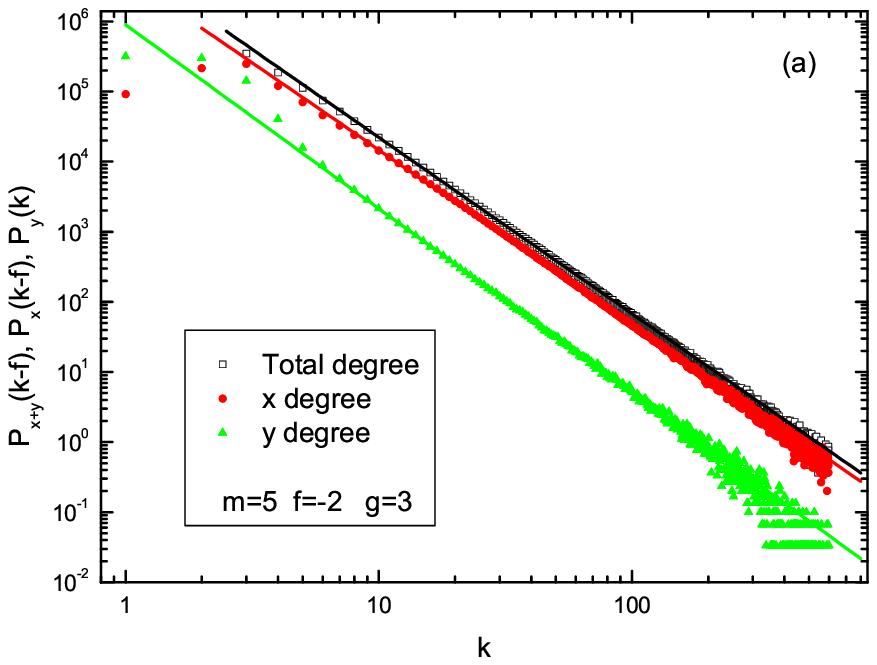,height=2.5in} \epsfig{file=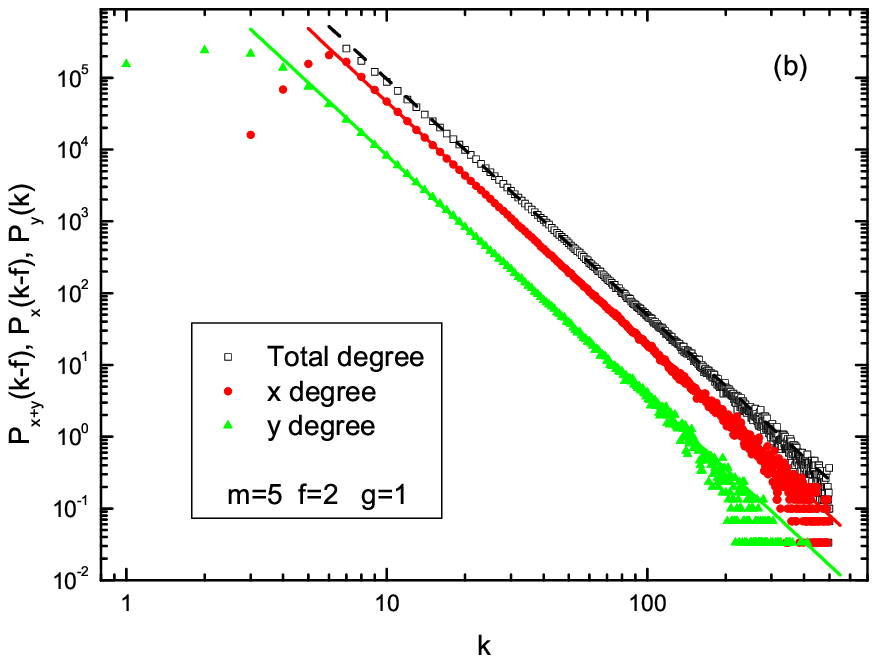,
height=2.5in} \epsfig{file=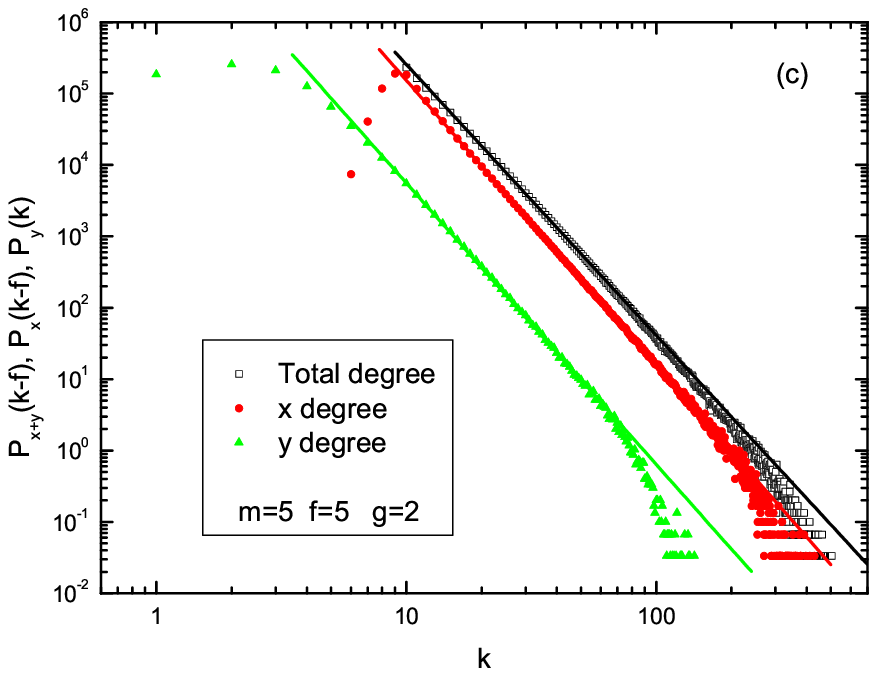,height=2.5in} \caption{The
distribution of total degree, $X$ degree and $Y$ degree. The size
of the network $N=1000000$ and $m=5$. The lines are linear fits of
the main
part of the data. (a) $f=-2$, $g=3$. The slopes of the lines are $\protect%
\gamma _{k}=-2.5$, $\protect\gamma _{x}=-2.5$ and $\protect\gamma %
_{y}=-2.6$. The prediction is $\protect\gamma =-2.6$. (b) $f=2$,
$g=1$. The
slopes of the lines are $\protect\gamma _{k}=-3.3$, $\protect\gamma %
_{x}=-3.4$ and $\protect\gamma _{y}=-3.4$. The prediction is $%
\protect\gamma =-3.4$. (a) $f=5$, $g=2$. The slopes of the lines are $%
\protect\gamma _{k}=-3.8$, $\protect\gamma _{x}=-4.0$ and $\protect%
\gamma _{y}=-3.9$. The prediction is $\protect\gamma =-4$.}
\label{figure1}
\end{figure}
\qquad

\begin{figure}[tbp]
\epsfig{file=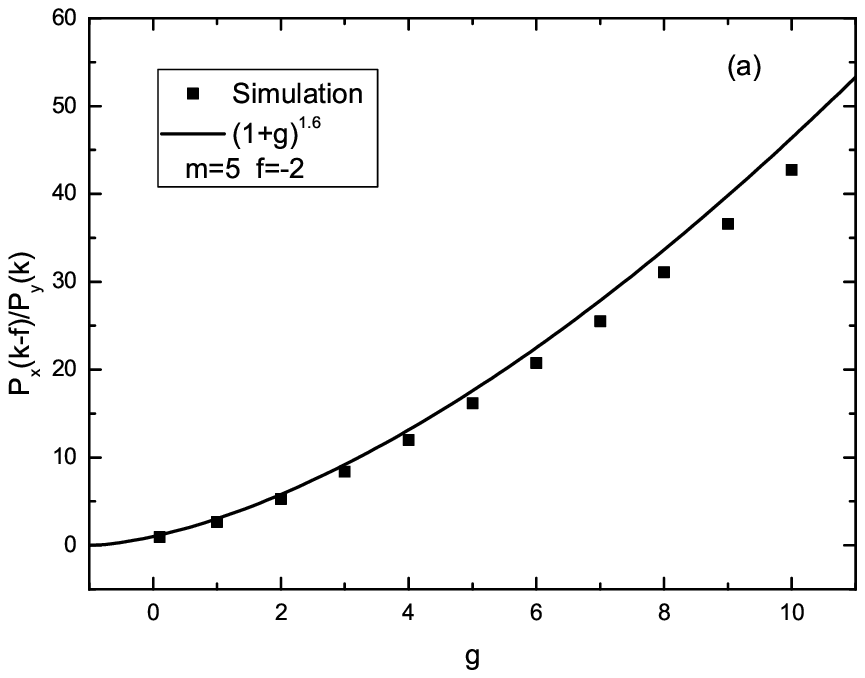,height=2.5in} %
\epsfig{file=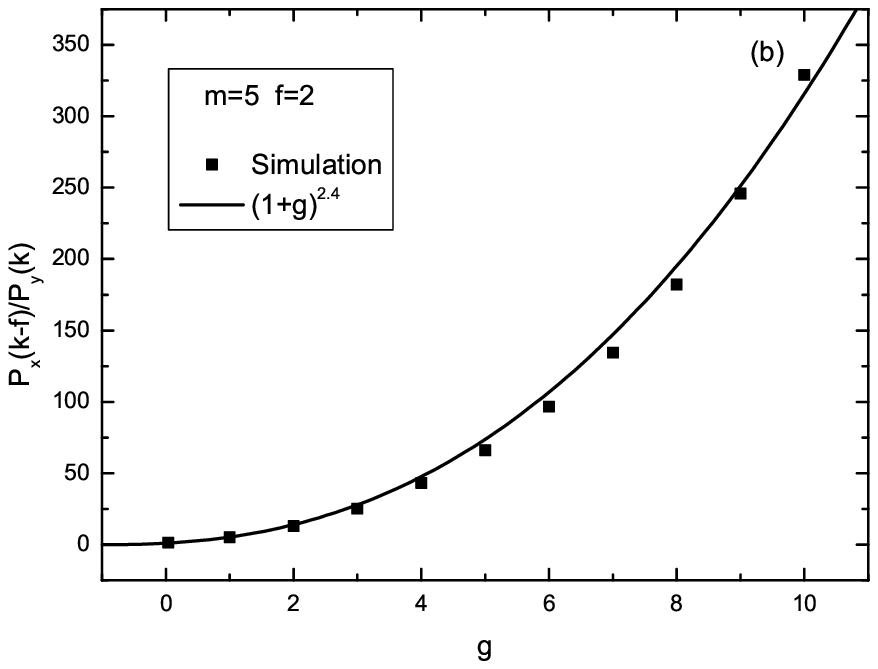,height=2.5in } %
\epsfig{file=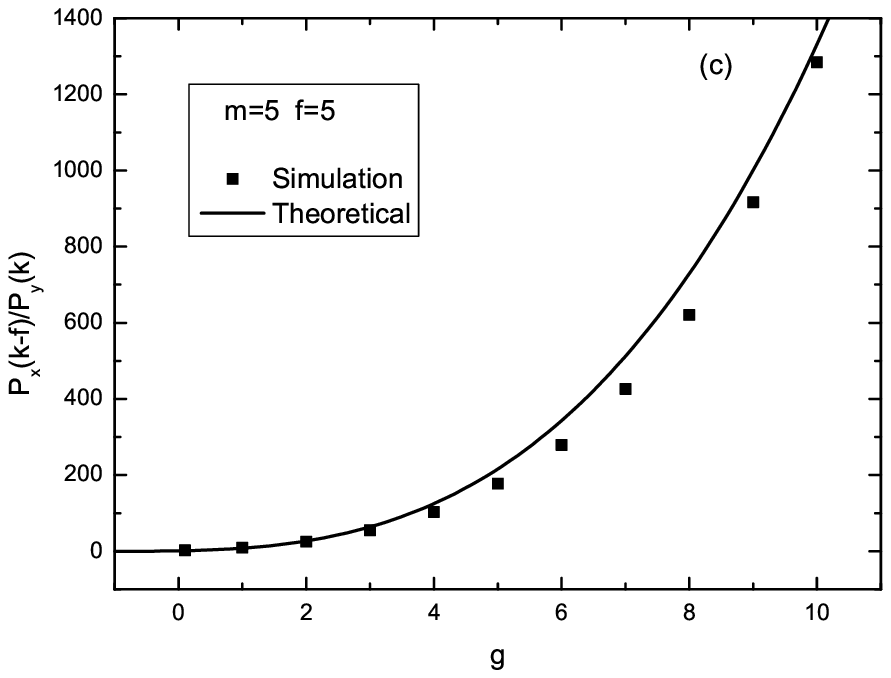,height=2.5in} \caption{The ratios of
the distribution $X$ degree to the distribution of $Y$ degree with
respective to different $g$. The curves are theoretical
predictions. The size of the network $N=1000000$ and $m=5$. (a) $f=-2$; (b) $%
f=2$; (c) $f=5$.}
\label{figure2}
\end{figure}

\begin{figure}[tbp]
\epsfig{file=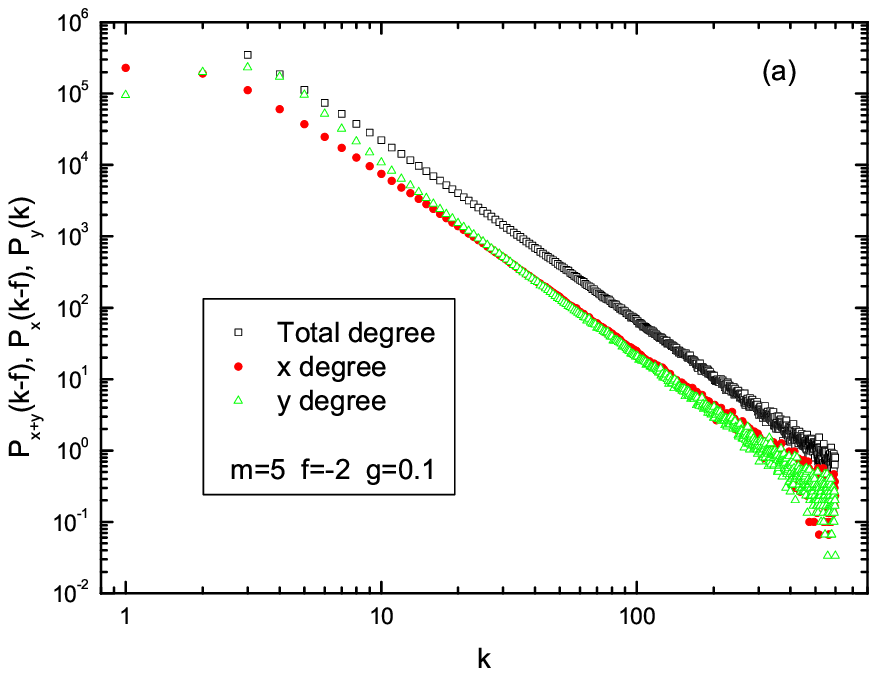,} \epsfig{file=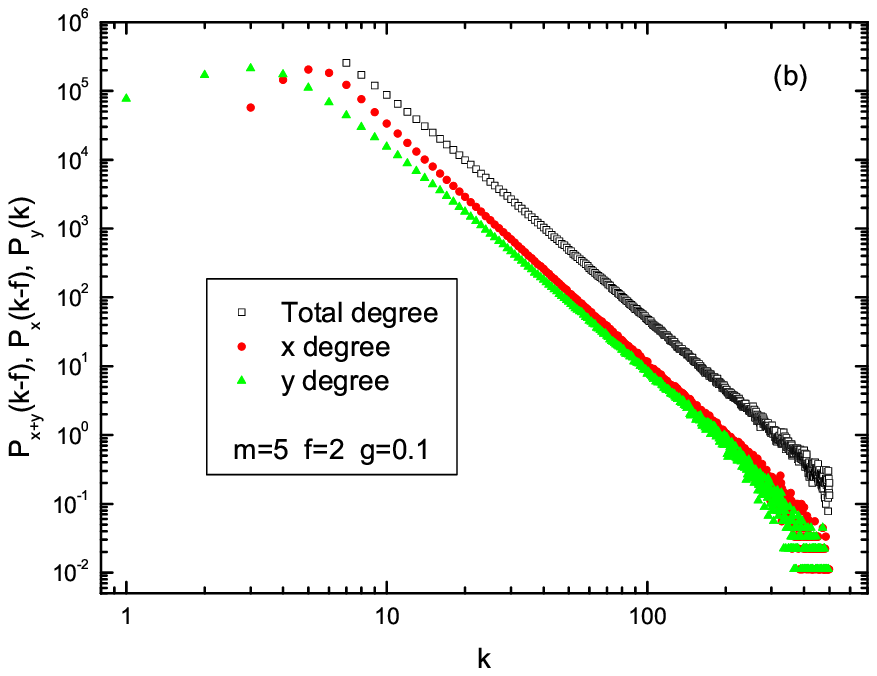,}
\caption{The degree distribution of total degree, $x$ degree and $y$ degree
with small $g=0.1$. The size of the network $N=1000000$ and $m=5$. (a) $f=-2$%
. (b) $f=2$.}
\label{figure3}
\end{figure}

\begin{figure}[tbp]
\epsfig{file=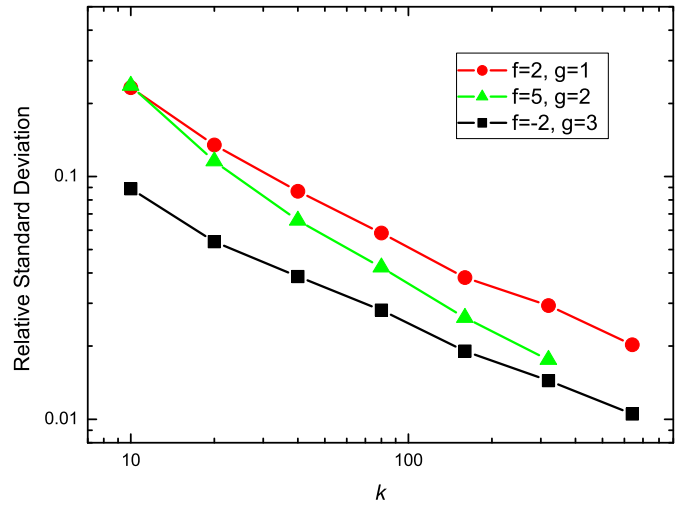,}\caption{The relative standard deviation of
$X$ degree respective to total degree. The size of the network
$N=1000000$.} \label{figure4}
\end{figure}
\end{document}